\documentstyle[twocolumn,aps,floats,epsfig]{revtex}

\begin{document}

\def\d{{\rm d}}
\def\e{{\rm e}}
\def\O{{\rm O}}
\def\half{\mbox{$\frac12$}}
\def\from{\leftarrow}
\def\eref#1{(\protect\ref{#1})}
\def\etal{{\it{}et~al.}}

\draft
\tolerance = 10000

\twocolumn[\hsize\textwidth\columnwidth\hsize\csname @twocolumnfalse\endcsname

\title{A fast Monte Carlo algorithm for site or bond percolation}
\author{M. E. J. Newman$^1$ and R. M. Ziff$^2$}
\address{$^1$Santa Fe Institute, 1399 Hyde Park Road, Santa Fe, NM 87501}
\address{$^2$Michigan Center for Theoretical Physics and Department
of Chemical Engineering,\\
University of Michigan, Ann Arbor, MI 48109}
\maketitle

\begin{abstract}
We describe in detail a new and highly efficient algorithm for studying
site or bond percolation on any lattice.  The algorithm can measure an
observable quantity in a percolation system for all values of the site or
bond occupation probability from zero to one in an amount of time which
scales linearly with the size of the system.  We demonstrate our algorithm
by using it to investigate a number of issues in percolation theory,
including the position of the percolation transition for site percolation
on the square lattice, the stretched exponential behavior of spanning
probabilities away from the critical point, and the size of the giant
component for site percolation on random graphs.
\end{abstract}

\vspace{1.0cm}

]

\section{Introduction}
Percolation~\cite{SA92} is certainly one of the best studied problems in
statistical physics.  Its statement is trivially simple: in site
percolation every site on a specified lattice is independently either
``occupied,'' with probability~$p$, or not with probability~$1-p$.  The
occupied sites form contiguous clusters which have some interesting
properties.  In particular, the system shows a continuous phase transition
at a finite value of $p$ which, on a regular lattice, is characterized by
the formation of a cluster large enough to span the entire system from one
side to the other in the limit of infinite system size.  We say such a
system ``percolates.''  As the phase transition is approached from small
values of $p$, the average cluster size diverges in a way reminiscent of
the divergence of fluctuations in the approach to a thermal continuous
phase transition, and indeed one can define correlation functions and a
correlation length in the obvious fashion for percolation models, and hence
measure critical exponents for the transition.

One can also consider bond percolation in which the bonds of the lattice
are occupied (or not) with probability $p$ (or $1-p$), and this system
shows behavior qualitatively similar to though different in some details
from site percolation.

Site and bond percolation have found a huge variety of uses in many fields.
Percolation models appeared originally in studies of actual percolation in
materials~\cite{SA92}---percolation of fluids through rock for
example~\cite{GG78,LSD81,Sahimi94}---but have since been used in studies of
many other systems, including granular materials~\cite{OT98,Tobochnik99},
composite materials~\cite{BFD92}, polymers~\cite{BHP95},
concrete~\cite{BG92}, aerogel and other porous media~\cite{Machta91,MG95},
and many others.  Percolation also finds uses outside of physics, where it
has been employed to model resistor networks~\cite{ARC85}, forest
fires~\cite{Henley93} and other ecological disturbances~\cite{WC95},
epidemics~\cite{NW99,MN00}, robustness of the Internet and other
networks~\cite{CEBH00,CNSW00}, biological evolution~\cite{RJ94}, and social
influence~\cite{SWAJS00}, amongst other things.  It is one of the simplest
and best understood examples of a phase transition in any system, and yet
there are many things about it that are still not known.  For example,
despite decades of effort, no exact solution of the site percolation
problem yet exists on the simplest two-dimensional lattice, the square
lattice, and no exact results are known on any lattice in three dimensions
or above.  Because of these and many other gaps in our current understanding
of percolation, numerical simulations have found wide use in the field.

Computer studies of percolation are simpler than most simulations in
statistical physics, since no Markov process or other importance sampling
mechanism is needed to generate states of the lattice with the correct
probability distribution.  We can generate states simply by occupying each
site or bond on an initially empty lattice with independent
probability~$p$.  Typically, we then want to find all contiguous clusters
of occupied sites or bonds, which can be done using a simple depth- or
breadth-first search.  Once we have found all the clusters, we can easily
calculate, for example, average cluster size, or look for a system spanning
cluster.  Both depth- and breadth-first searches take time $\O(M)$ to find
all clusters, where $M$ is the number of bonds on the lattice.  This is
optimal, since one cannot check the status of $M$ individual bonds or
adjacent pairs of sites in any less than $\O(M)$ operations~\cite{note1}.
On a regular lattice, for which $M=\half z N$, where $z$ is the
coordination number and $N$ is the number of sites, $\O(M)$ is also
equivalent to $\O(N)$.

But now consider what happens if, as is often the case, we want to
calculate the values of some observable quantity $Q$ (e.g.,~average cluster
size) over a range of values of~$p$.  Now we have to perform repeated
simulations at many different closely-spaced values of $p$ in the range of
interest, which makes the calculation much slower.  Furthermore, if we want
a continuous curve of $Q(p)$ in the range of interest then in theory we
have to measure $Q$ at an infinite number of values of~$p$.  More
practically, we can measure it at a finite number of values and then
interpolate between them, but this inevitably introduces some error into
the results.

The latter problem can be solved easily enough.  The trick~\cite{Hu92,GT96}
is to measure $Q$ for {\em fixed\/} numbers of occupied sites (or bonds)
$n$ in the range of interest.  Let us refer to the ensemble of states of a
percolation system with exactly $n$ occupied sites or bonds as a
``microcanonical percolation ensemble,'' the number $n$ playing the role of
the energy in thermal statistical mechanics.  The more normal case in which
only the occupation probability~$p$ is fixed is then the ``canonical
ensemble'' for the problem.  (If one imagines the occupied sites or bonds
as representing particles instead, then the two ensembles would be
``canonical'' and ``grand canonical,'' respectively.  Some
authors~\cite{SV00} have used this nomenclature.)  Taking site percolation
as an example, the probability of there being exactly $n$ occupied sites on
the lattice for a canonical percolation ensemble is given by the binomial
distribution:
\begin{equation}
B(N,n,p) = \biggl({N\atop n}\biggr) p^n (1-p)^{N-n}.
\label{binomial}
\end{equation}
(The same expression applies for bond percolation, but with $N$ replaced by
$M$, the total number of bonds.)  Thus, if we can measure our observable
within the microcanonical ensemble for all values of $n$, giving a set of
measurements $\lbrace Q_n\rbrace$, then the value in the canonical ensemble
will be given by
\begin{equation}
Q(p) = \sum_{n=0}^N B(N,n,p) Q_n 
     = \sum_{n=0}^N \biggl({N\atop n}\biggr) p^n (1-p)^{N-n} Q_n.
\label{convolution}
\end{equation}
Thus we need only measure $Q_n$ for all values of $n$, of which there are
$N+1$, in order to find $Q(p)$ for all~$p$.  Since filling the lattice
takes time $\O(N)$, and construction of clusters $\O(M)$, we take time
$\O(M+N)$ for each value of $n$, and $\O(N^2+MN)$ for the entire
calculation, or more simply $\O(N^2)$ on a regular lattice.  Similar
considerations apply for bond percolation.

In fact, one frequently does not want to know the value of $Q(p)$ over the
entire range $0\le p\le1$, but only in the critical region, i.e.,~the
region in which the correlation length $\xi$ is greater than the linear
dimension $L$ of the system.  If $\xi\sim|p-p_c|^{-\nu}$ close to $p_c$,
where $\nu$ is a (constant) critical exponent, then the width of the
critical region scales as $L^{-1/\nu}$.  Since the binomial distribution,
Eq.~\eref{binomial}, falls off very fast away from its maximum
(approximately as a Gaussian with variance of order $N^{-1}$), it is safe
also only to calculate $Q_n$ within a region scaling as $L^{-1/\nu}$, and
there are order $L^{d-1/\nu}$ values of $N$ in such a region, where $d$ is
the dimensionality of the lattice.  Thus the calculation of $Q(p)$ in the
critical region will take time of order $N L^{d-1/\nu} = N^{2-1/d\nu}$.  In
two dimensions, for example, $\nu$ is known to take the value $\frac43$,
and so we can calculate any observable quantity over the whole critical
region in time~$\O(N^{13/8})$.

An alternative technique for calculating $Q(p)$ over a range of values of
$p$ is to use a histogram method.  (This should not be confused with the
method of Hu~\cite{Hu92}, which is also referred to as a ``histogram
method,'' but which is different from the method described here.  Our
histogram method is the equivalent for a percolation model of the method of
Ferrenberg and Swendsen~\cite{FS88} for thermal models.)  Suppose that one
performs a number $M$ of simulations of a percolation system, each one
generating a single state $\mu$ of the system drawn from some probability
distribution~$P_\mu$.  In each state we measure our observable of interest
$Q$, giving a set of measurements $\lbrace Q_\mu\rbrace$.  Then our
estimate of the value of $Q(p)$ is given by
\begin{equation}
Q(p) = {\sum_\mu P_\mu^{-1} B(N,n_\mu,p) Q_\mu\over
        \sum_\mu P_\mu^{-1} B(N,n_\mu,p)},
\label{histogram}
\end{equation}
where $n_\mu$ is the number of occupied sites or bonds in state~$\mu$.  In
normal (canonical) simulations of percolation $P_\mu$ is just equal to the
binomial distribution $B(N,n_\mu,p)$, which then cancels out of
Eq.~\eref{histogram} to give $Q(p)={1\over M} \sum_\mu Q_\mu$ in the usual
fashion.  In microcanonical simulations $P_\mu$ is uniform over values of
$n$, and Eq.~\eref{histogram} is equivalent to Eq.~\eref{convolution}.
Suppose however that instead $P_\mu$ corresponds to the distribution of
states produced by a canonical simulation at a different value $p_0$ of the
site or bond occupation probability.  In this case $P_\mu=B(N,n_\mu,p_0)$,
and Eq.~\eref{histogram} becomes
\begin{eqnarray}
Q(p) &=& {\sum_\mu [B(N,n_\mu,p)/B(N,n_\mu,p_0)] Q_\mu\over
          \sum_\mu  B(N,n_\mu,p)/B(N,n_\mu,p_0)}\nonumber\\
     &=& {\sum_\mu {\bigl[\bigl({1\over p_0}-1\bigr)\big/
                  \bigl({1\over p}-1\bigr)\bigr]}^{n_\mu} Q_\mu\over
          \sum_\mu {\bigl[\bigl({1\over p_0}-1\bigr)\big/
                  \bigl({1\over p}-1\bigr)\bigr]}^{n_\mu}}.
\end{eqnarray}
This equation tells us that if we perform a simulation or set of
simulations at occupation probability $p_0$, and record the value(s) of
$n_\mu$ and $Q_\mu$ for each state generated, we can use the results to
estimate $Q(p)$ at any other value of~$p$.  In practice, the range over
which we can reliably extrapolate away from $p_0$ using this method is
limited by the range of values of $n$ sampled; if the values of $n$ sampled
at $p_0$ make negligible contributions to the canonical ensemble of states
at $p$, then extrapolation will give poor results.  Since the values
$n_\mu$ are drawn from a binomial distribution with width
$\sqrt{Np_0(1-p_0)}$, the entire range of $n$ can be sampled with
$\O(N^{1/2})$ separate simulations at different values of~$p_0$.  Each
simulation takes time $\O(N)$ to complete, and hence it takes time
$\O(N^{3/2})$ to calculate $Q(p)$ for all values of~$p$.  If we are only
interested in the critical region, this time is reduced to
$\O(N^{3/2-1/d\nu})$, which is $\O(N^{9/8})$ in two dimensions.  These
times are considerably faster than those for the direct method described
above of performing simulations for all values of $n$, but this speed is
offset by the fact that the histogram method gives larger statistical
errors on measured quantities than the direct method~\cite{FLS95,NP99}.

In this paper we describe a new algorithm which can perform the calculation
of $Q(p)$ for all values of $p$ faster than any of the algorithms above, in
time $\O(N)$.  This is clearly far superior to the $\O(N^2)$ direct
algorithm while having comparable accuracy in terms of statistical error.
It is also substantially faster than the $\O(N^{3/2})$ histogram method
while having significantly better statistical accuracy.  And even in the
case where we only want to calculate $Q$ in the critical region, our
algorithm is still faster doing all values of $n$, than previous algorithms
doing only those in the critical region.  (Because of the way it works, our
algorithm cannot be used to calculate $Q_n$ only for the critical region;
one is obliged to calculate $\O(N)$ values of $Q_n$.)  A typical lattice
size for percolation systems is about a million sites.  On such a lattice
our algorithm would be on the order of a million times faster than the
simple $\O(N^2)$ method and a thousand times faster than the histogram
method, and even just within the critical region it would still be around
six times faster than the histogram method in two dimensions, while giving
substantially better statistical accuracy.

The outline of this paper is as follows.  In Section~\ref{algorithm} we
describe our algorithm.  In Section~\ref{applications} we give results from
the application of the algorithm to three different example problems.  In
Section~\ref{concs} we give our conclusions.  We have reported some of the
results given here previously in Ref.~\onlinecite{NZ00}.

\section{The algorithm}
\label{algorithm}
Our algorithm is based on a very simple idea.  In the standard algorithms
for percolation, one must create an entire new state of the lattice for
every different value of $n$ one wants to investigate, and construct the
clusters for that state.  As various authors have pointed out,
however~\cite{Hu92,GT96,Moukarzel98,FLR99}, if we want to generate states
for each value of $n$ from zero up to some maximum value, then we can save
ourselves some effort by noticing that a correct sample state with $n+1$
occupied sites or bonds is given by adding one extra randomly chosen site
or bond to a correct sample state with $n$ sites or bonds.  In other words,
we can create an entire set of correct percolation states by adding sites
or bonds one by one to the lattice, starting with an empty lattice.

Furthermore, the configuration of clusters on the lattice changes little
when only a single site or bond is added, so that, with a little ingenuity,
one can calculate the new clusters from the old with only a small amount of
computational effort.  This is the idea at the heart of our algorithm.

Let us consider the case of bond percolation first, for which our algorithm
is slightly simpler.  We start with a lattice in which all $M$ bonds are
unoccupied, so that each of the $N$ sites is its own cluster with just one
element.  As bonds are added to the lattice, these clusters will be joined
together into larger ones.  The first step in our algorithm is to decide an
order in which the bonds will be occupied.  We wish to choose this order
uniformly at random from all possible such orders, i.e.,~we wish to choose
a random permutation of the bonds.  One way of achieving this is the
following:
\begin{enumerate}
\item Create a list of all the bonds in any convenient order.  Positions in
this list are numbered from $1$ to~$M$.
\item Set $i\from1$.
\item Choose a number $j$ uniformly at random in the range $i\le j\le M$.
\item Exchange the bonds in positions $i$ and $j$.  (If $i=j$ then nothing
happens.)
\item Set $i\from i+1$.
\item Repeat from step~3 until $i=M$.
\end{enumerate}
It is straightforward to convince oneself that all permutations of the
bonds are generated by this procedure with equal probability, in time
$\O(M)$, i.e.,~going linearly with the number of bonds on the lattice.

\begin{figure}
\begin{center}
\psfig{figure=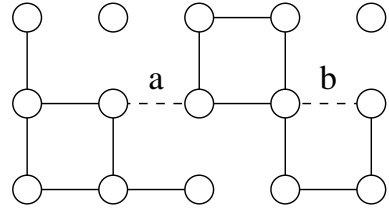,width=5cm}
\end{center}
\caption{Two examples of bonds (dotted lines) being added to
bond-percolation configurations.  (a)~The added bond joins two clusters
together.  (b)~The added bond does nothing, since the two joined sites were
already members of the same cluster.}
\label{joining}
\end{figure}

Having chosen an order for our bonds, we start to occupy them in that
order.  The first bond added will, clearly, join together two of our
single-site clusters to form a cluster of two sites, as will, almost
certainly, the second and third.  However, not all bonds added will join
together two clusters, since some bonds will join pairs of sites which are
already part of the same cluster---see Fig.~\ref{joining}.  Thus, in order
to correctly keep track of the cluster configuration of the lattice, we
must do two things for each bond added:

{\bf Find:} when a bond is added to the lattice we must find which
clusters the sites at either end belong to.

{\bf Union:} if the two sites belong to different clusters, those
clusters must be amalgamated into a single cluster; otherwise, if the two
belong to the same cluster, we need do nothing.

Algorithms which achieve these steps are known as ``union/find''
algorithms~\cite{Sedgewick88,Knuth97}.  Union/find algorithms are widely
used in data structures, in calculations on graphs and trees, and in
compilers.  They have been extensively studied in computer science, and we
can make profitable use of a number of results from the computer science
literature to implement our percolation algorithm simply and efficiently.

It is worth noting that measured values for lattice quantities such as,
say, average cluster size, are not statistically independent in our method,
since the configuration of the lattice changes at only one site from one
step of the algorithm to the next.  This contrasts with the standard
algorithms, in which a complete new configuration is generated for every
value of $n$ or $p$ investigated, and hence all data points are
statistically independent.  While this is in some respects a disadvantage
of our method, it turns out that in most cases of interest it is not a
problem, since almost always one is concerned only with statistical
independence of the results from one {\em run\/} to another.  This our
algorithm clearly has, which means that an error estimate on the results
made by calculating the variance over many runs will be a correct error
estimate, even though the errors on the observed quantity for successive
values of $n$ or $p$ will not be independent.

In the next two sections we apply a number of different and increasingly
efficient union/find algorithms to the percolation problem, culminating
with a beautifully simple and almost linear algorithm associated with the
names of Michael Fischer and Robert Tarjan.

\begin{figure}
\begin{center}
\psfig{figure=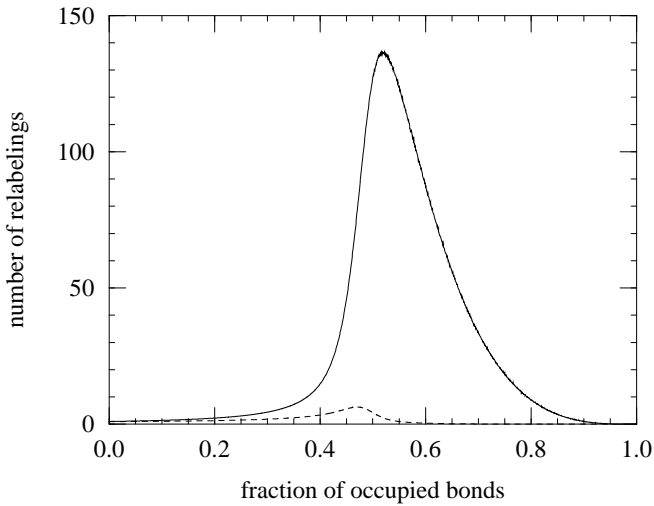,width=\columnwidth}
\end{center}
\caption{Number of relabelings of sites performed, per bond added, as a
function of fraction of occupied bonds, for the two algorithms described in
Section~\ref{trivial}, on a square lattice of $32\times32$ sites.}
\label{csd}
\end{figure}

\subsection{Trivial (but possibly quite efficient)\\
union/find algorithms}
\label{trivial}
Perhaps the simplest union/find algorithm which we can employ for our
percolation problem is the following.  We add a label to each site of our
lattice---an integer for example---which tells us to which cluster that
site belongs.  Initially, all such labels are different (e.g.,~they could
be set equal to their site label).  Now the ``find'' portion of the
algorithm is simple---we examine the labels of the sites at each end of an
added bond to see if they are the same.  If they are not, then the sites
belong to different clusters, which must be amalgamated into one as a
result of the addition of this bond.  The amalgamation, the ``union''
portion of the algorithm, is more involved.  To amalgamate two clusters we
have to choose one of them---in the simplest case we just choose one at
random---and set the cluster labels of all sites in that cluster equal to
the cluster label of the other cluster.

In the initial stages of the algorithm, when most bonds are unoccupied,
this amalgamation step will be quick and simple; most bonds added will
amalgamate clusters, but the clusters will be small and only a few sites
will have to be relabeled.  In the late stages of the algorithm, when most
bonds are occupied, all or most sites will belong to the system-size
percolating cluster, and hence cluster unions will rarely be needed, and
again the algorithm is quick.  Only in the intermediate regime, close to
the percolation point, will any significant amount of work be required.  In
this region, there will in general be many large clusters, and much
relabeling may have to be performed when two clusters are amalgamated.
Thus, the algorithm displays a form of critical slowing down as it passes
through the critical region.  We illustrate this in Fig.~\ref{csd} (solid
line), where we show the number of site relabelings taking place, as a
function of fraction of occupied bonds, for bond percolation on a square
lattice of $32\times32$ sites.  The results are averaged over a million
runs of the algorithm, and show a clear peak in the number of relabelings
in the region of the known percolation transition at bond occupation
probability $p_c=\half$.

\begin{figure}
\begin{center}
\psfig{figure=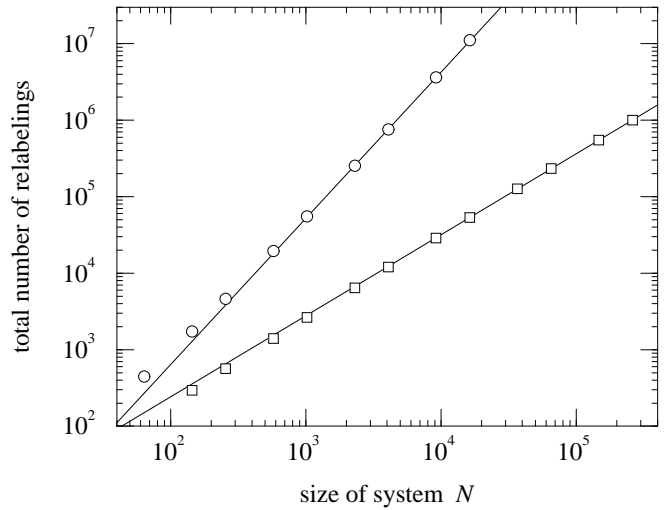,width=\columnwidth}
\end{center}
\caption{Total number of relabelings of sites performed in a single run as
a function of system size for the two algorithms (circles and squares) of
Section~\ref{trivial}.  Each data point is an average over 1000 runs; error
bars are much smaller than the points themselves.  The solid lines are
straight-line fits to the rightmost five points in each case.}
\label{simplescale}
\end{figure}

In Fig.~\ref{simplescale} (circles) we show the average total number of
relabelings which are carried out during the entire run of the algorithm on
square lattices of $N=L\times L$ sites as a function of~$N$.  Given that,
as the number of relabelings becomes large, the time taken to relabel will
be the dominant factor in the speed of this algorithm, this graph gives an
indication of the expected scaling of run-time with system size.  As we can
see, the results lie approximately on a straight line on the logarithmic
scales used in the plot and a fit to this line gives a run-time which
scales as $N^\alpha$, with $\alpha=1.91\pm0.01$.  This is slightly better
than the $\O(N^2)$ behavior of the standard algorithms, and in practice can
make a significant difference to running time on large lattices.  With a
little ingenuity however, we can do a lot better.

The algorithm's efficiency can be improved considerably by one minor
modification: we maintain a separate list of the sizes of the clusters,
indexed by their cluster label, and when two clusters are amalgamated,
instead of relabeling one of them at random, we look up their sizes in this
list and then relabel the smaller of the two.  This ``weighted union''
strategy ensures that we carry out the minimum possible number of
relabelings, and in particular avoids repeated relabelings of the large
percolating cluster when we are above the percolation transition, which is
a costly operation.  The average number of relabelings performed in this
algorithm as a function of number of occupied bonds is shown as the dotted
line in Fig.~\ref{csd}, and it is clear that this change in the algorithm
has saved us a great deal of work.  In Fig.~\ref{simplescale} (squares) we
show the total number of relabelings as a function of system size, and
again performance is clearly much better.  The run-time of the algorithm
now appears to scale as $N^\alpha$ with $\alpha=1.06\pm0.01$.  In fact, we
can prove that the worst-case running time goes as $N\log N$, a form which
frequently manifests itself as an apparent power law with exponent slightly
above~1.  To see this consider that with weighted union the size of the
cluster to which a site belongs must at least double every time that site
is relabeled.  The maximum number of such doublings is limited by the size
of the lattice to $\log_2 N$, and hence the maximum number of relabelings
of a site has the same limit.  An upper bound on the total number of
relabelings performed for all $N$ sites during the course of the algorithm
is therefore $N\log_2 N$.  (A similar argument applied to the unweighted
relabeling algorithm implies that $N^2$ is an upper bound on the running
time of this algorithm.  The measured exponent $1.91$ indicates either that
the experiments we have performed have not reached the asymptotic scaling
regime, or else that the worst-case running time is not realized for the
particular case of union of percolation clusters on a square lattice.)

An algorithm whose running time scales as $\O(N\log N)$ is a huge advance
over the standard methods.  However, it turns out that, by making use of
some known techniques and results from computer science, we can make our
algorithm better still while at the same time actually making it simpler.
This delightful development is described in the next section.

\begin{figure}
\begin{center}
\psfig{figure=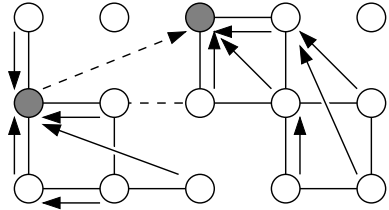,width=5cm}
\end{center}
\caption{An example of the tree structure described in the text, here
  representing two clusters.  The shaded sites are the root sites and the
  arrows represent pointers.  When a bond is added (dotted line in center)
  that joins sites which belong to different clusters (i.e.,~whose pointers
  lead to different root sites), the two clusters must be amalgamated by
  making one (left) a subtree of the other (right).  This is achieved by
  adding a new pointer from the root of one tree to the root of the other.}
\label{trees}
\end{figure}

\subsection{Tree-based union/find algorithms}
Almost all modern union/find algorithms make use of data trees to store the
sets of objects (in this case clusters) which are to be searched.  The idea
of using a tree in this way seems to have been suggested first by Galler
and Fischer~\cite{GF64}.  Each cluster is stored as a separate tree, with
each vertex of the tree being a separate site in the cluster.  Each cluster
has a single ``root'' site, which is the root of the corresponding tree,
and all other sites possess pointers either to that root site or to another
site in the cluster, such that by following a succession of such pointers
we can get from any site to the root.  By traversing trees in this way, it
is simple to ascertain whether two sites are members of the same cluster:
if their pointers lead to the same root site then they are, otherwise they
are not.  This scheme is illustrated for the case of bond percolation on
the square lattice in Fig.~\ref{trees}.

The union operation is also simple for clusters stored as trees: two
clusters can be amalgamated simply by adding a pointer from the root of one
to any site in the other (Fig.~\ref{trees}).  Normally, one chooses the new
pointer to point to the root of the other cluster, since this reduces the
average distance that will be have to be traversed across the tree to reach
the root site from other sites.

There are many varieties of tree-based union/find algorithms, which differ
from one another in the details of how the trees are updated as the
algorithm progresses.  However, the best known performance for any such
algorithm is for a very simple one---the ``weighted union/find with path
compression''---which was first proposed by
Fischer~\cite{Sedgewick88,Knuth97,Fischer72}.  Its description is brief:

{\bf Find:} In the find part of the algorithm, trees are traversed to find
their root sites.  If two initial sites lead to the same root, then they
belong to the same cluster.  In addition, after the traversal is completed,
all pointers along the path traversed are changed to point directly to the
root of their tree.  This is called ``path compression.''  Path compression
makes traversal faster the next time we perform a find operation on any of
the same sites.

{\bf Union:} In the union part of the algorithm, two clusters are
amalgamated by adding a pointer from the root of one to the root of the
other, thus making one a subtree of the other.  We do this in a
``weighted'' fashion as in Section~\ref{trivial}, meaning that we always
make the smaller of the two a subtree of the larger.  In order to do this,
we keep a record at the root site of each tree of the number of sites in
the corresponding cluster.  When two clusters are amalgamated the size of
the new composite cluster is the sum of the sizes of the two from which it
was formed.

Thus our complete percolation algorithm can be summarized as follows.
\begin{enumerate}
\item Initially all sites are clusters in their own right.  Each is its own
root site, and contains a record of its own size, which is~1.
\item Bonds are occupied in random order on the lattice.
\item Each bond added joins together two sites.  We follow pointers from
  each of these sites separately until we reach the root sites of the
  clusters to which they belong.  Root sites are identified by the fact
  that they do not possess pointers to any other sites.  Then we go back
  along the paths we followed through each tree and adjust all pointers
  along those paths to point directly to the corresponding root sites.
\item If the two root sites are the same site, we need do nothing further.
\item If the two root sites are different, we examine the cluster sizes
  stored in them, and add a pointer from the root of the smaller cluster to
  the root of the larger, thereby making the smaller tree a subtree of the
  larger one.  If the two are the same size, we may choose whichever tree
  we like to be the subtree of the other.  We also update the size of the
  larger cluster by adding the size of the smaller one to it.
\end{enumerate}
These steps are repeated until all bonds on the lattice have been occupied.
At each step during the run, the tree structures on the lattice correctly
describe all of the clusters of joined sites, allowing us to evaluate
observable quantities of interest.  For example, if we are interested in
the size of the largest cluster on the lattice as a function of the number
of occupied bonds, we simply keep track of the largest cluster size we have
seen during the course of the algorithm.

Although this algorithm may seem more involved than the relabeling
algorithm of Section~\ref{trivial}, it turns out that its implementation is
actually simpler, for two reasons.  First, there is no relabeling of sites,
which eliminates the need for the code portion which tracks down all the
sites belonging to a cluster and updates them.  Second, the only difficult
parts of the algorithm, the tree traversal and path compression, can it
turns out be accomplished together by a single function which, by artful
use of recursion, can be coded in only three lines (two in~C).  The
implementation of the algorithm is discussed in detail in
Section~\ref{implementation} and in Appendix~A.

\subsection{Performance of the algorithm}
Without either weighting or path compression each step of the algorithm
above is known to take time linear in the tree size~\cite{Fischer72}, while
with either weighting or path compression alone it takes logarithmic
time~\cite{Fischer72,Tarjan75}.  When both are used together, however, each
step takes an amount of time which is {\em very nearly\/} constant, in a
sense which we now explain.

The worst-case performance of the weighted union/find algorithm with path
compression has been analyzed by Tarjan~\cite{Tarjan75} in the general case
in which one starts with $n$ individual single-element sets and performs
all $n-1$ unions (each one reducing the number of sets by 1 from $n$ down
to~1), and any number $m\ge n$ of intermediate finds.  The finds are
performed on any sequence of items, and the unions may be performed in any
order.  Our case is more restricted.  We always perform exactly $2M$ finds,
and only certain unions and certain finds may be performed.  For example,
we only ever perform finds on pairs of sites which are adjacent on the
lattice.  And unions can only be performed on clusters which are adjacent
to one another along at least one bond.  However, it is clear that the
worst-case performance of the algorithm in the most general case also
provides a bound on the performance of the algorithm in any more restricted
case, and hence Tarjan's result applies to our algorithm.

Tarjan's analysis is long and complicated.  However, the end result is
moderately straightforward.  The union part of the algorithm clearly takes
constant time.  The find part takes time which scales as the number of
steps taken to traverse the tree to its root.  Tarjan showed that the
average number of steps is bounded above by $k\alpha(m,n)$, where $k$ is an
unknown constant and the function $\alpha(m,n)$ is the smallest integer
value of $z$ greater than or equal to~1, such that $A(z,4\lceil m/n
\rceil)>\log_2 n$.  The function $A(i,j)$ in this expression is a slight
variant on Ackermann's function~\cite{Ackermann28} defined by
\begin{eqnarray}
A(0,j) &=& 2j \hspace{33.6mm}\mbox{for all $j$}\nonumber\\
A(i,0) &=& 0  \hspace{35.3mm}\mbox{for $i\ge1$}\nonumber\\
A(i,1) &=& 2  \hspace{35.3mm}\mbox{for $i\ge1$}\nonumber\\
A(i,j) &=& A(i-1, A(i,j-1)) \qquad\mbox{for $i\ge1$, $j\ge2$.}
\label{ackermann}
\end{eqnarray}
This function is a monstrously quickly growing function of its
arguments---faster than any multiple exponential---which means that
$\alpha(m,n)$, which is roughly speaking the functional inverse of
$A(i,j)$, is a very slowly growing function of its arguments.  So slowly
growing, in fact, that for all practical purposes it is a constant.  In
particular, note that $\alpha(m,n)$ is maximized by setting $m=n$, in which
case if $\log_2 n<A(z,4)$ then $\alpha(m,n)\le\alpha(n,n)\le z$.  Setting
$z=3$ and making use of Eq.~\eref{ackermann}, we find that
\begin{equation}
A(3,4) = {{{2^2}^2}^{\cdots}}^2 \Bigr\rbrace\:\mbox{$65\,536$ twos}.
\end{equation}
This number is ludicrously large.  For example, with only the first 5 out
of the $65\,536$ twos, we have
\begin{equation}
{{{2^2}^2}^2}^2 = 2.0\times10^{19728}.
\end{equation}
This number is far, far larger than the number of atoms in the known
universe.  Indeed, a generous estimate of the number of Planck volumes in
the universe would put it at only around $10^{200}$.  So
it is entirely safe to say that $\log_2 n$ will never exceed $A(3,4)$,
bearing in mind that $n$ in our case is equal to $N$, the number of sites
on the lattice.  Hence $\alpha(m,n)<3$ for all practical values of $m$ and
$n$ and certainly for all lattices which fit within our universe.  Thus the
average number of steps needed to traverse the tree is at least one and at
most $3k$, where $k$ is an unknown constant.  The value of $k$ is found
experimentally to be of order~1.

What does this mean for our percolation algorithm?  It means that the time
taken by both the union and find steps is $\O(1)$ in the system size, and
hence that each bond added to the lattice takes time $\O(1)$.  Thus it
takes time $\O(M)$ to add all $M$ bonds, while maintaining an up-to-date
list of the clusters at all times.  (To be precise, the time to add all $M$
bonds is bounded above by $\O(M\alpha(2M,N))$, where $\alpha(2M,N)$ is not
expected to exceed~3 this side of kingdom come.)

\begin{figure}
\begin{center}
\psfig{figure=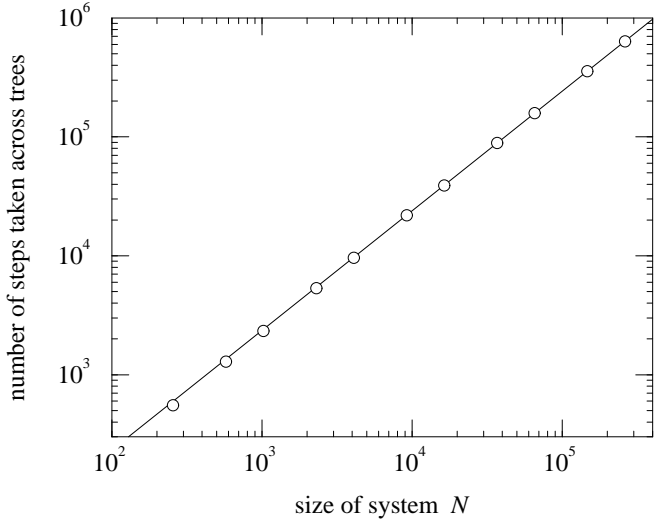,width=\columnwidth}
\end{center}
\caption{Total number of steps taken through trees during a single run of
the our algorithm as a function of system size.  Each point is averaged
over 1000 runs.  Statistical errors are much smaller than the data points.
The solid line is a straight-line fit to the last five points, and gives a
slope of $1.006\pm0.011$.}
\label{example}
\end{figure}

In Fig.~\ref{example} we show the actual total number of steps taken in
traversing trees during the entire course of a run of the algorithm
averaged over several runs for each data point, as a function of system
size.  A straight-line fit to the data shows that the algorithm runs in
time proportional to $N^\alpha$, with $\alpha=1.006\pm0.011$.  (In actual
fact, if one measures the performance of the algorithm in real
(``wallclock'') time, it will on most computers (circa~2001) not be
precisely linear in system size because of the increasing incidence of
cache misses as $N$ becomes large.  This however is a hardware failing,
rather than a problem with the algorithm.)

\begin{table}
\setlength{\tabcolsep}{6pt}
\begin{center}
\begin{tabular}{clcc}
 & algorithm & time in seconds & \\
\hline
 & depth-first search    & $4\,500\,000$ & \\
 & unweighted relabeling & $16\,000$     & \\
 & weighted relabeling   & $4.2$         & \\
 & tree-based algorithm  & $2.9$         & \\
\end{tabular}
\end{center}
\caption{Time in seconds for a single run of each of the algorithms
discussed in this paper, for bond percolation on a square lattice of
$1000\times1000$ sites.}
\label{timings}
\end{table}

We illustrate the comparative superiority of performance of our algorithm
in Table~\ref{timings} with actual timing results for it and the three
other algorithms described in this paper, for square lattices of
$1000\times1000$ sites, a typical size for numerical studies of
percolation.  All programs were compiled using the same compiler and run on
the same computer.  As the table shows, our algorithm, in addition to being
the simplest to program, is also easily the fastest.  In particular we find
it to be about 2~million times faster than the simple depth-first
search~\cite{note2}, and about 50\% faster than the best of our relabeling
algorithms, the weighted relabeling of Section~\ref{trivial}.  For larger
systems the difference will be greater still.

Site percolation can be handled by only a very slight modification of the
algorithm.  In this case, sites are occupied in random order on the
lattice, and each site occupied is declared to be a new cluster of size~1.
Then one adds, one by one, all bonds between this site and the occupied
sites, if any, adjacent to it, using the algorithm described above.  The
same performance arguments apply: generation of a permutation of the sites
takes time $\O(N)$, and the adding of all the bonds takes time $\O(M)$, and
hence the entire algorithm takes time $\O(M+N)$.  On a regular lattice, for
which $M=\half zN$, where $z$ is the coordination number, this is
equivalent to $\O(N)$.

It is worth also mentioning the memory requirements of our algorithm.  In
its simplest form, the algorithm requires two arrays of size $N$ for its
operation; one is used to store the order in which sites will be occupied
(for bond percolation this array is size~$M$), and one to store the
pointers and cluster sizes.  The standard depth-first search also requires
two arrays of size $N$ (a cluster label array and a stack), as does the
weighted relabeling algorithm (a label array and a list of cluster sizes).
Thus all the algorithms are competitive in terms of memory use.  (The
well-known Hoshen--Kopelman algorithm~\cite{HK76}, which is a variation on
depth-first search and runs in $\O(N^2\log N)$ time, has significantly
lower memory requirements, of order $N^{1-1/d}$, which makes it useful for
studies of particularly large low-dimensional systems.)

\subsection{Implementation}
\label{implementation}
In Appendix~A we give a complete computer program for our algorithm for
site percolation on the square lattice written in~C.  The program is also
available for download on the Internet~\cite{download}.  As the reader will
see, the implementation of the algorithm itself is quite simple, taking
only a few lines of code.  In this section, we make a few points about the
algorithm's implementation, which may be helpful to those wishing to make
use of it.

First, we note that the ``find'' operation in which a tree is traversed to
find its root and the path compression can be conveniently performed
together using a recursive function which in pseudocode would look like
this:
\begin{tabbing}
xxx\=xxx\=\kill
\> {\bf function} {\it find\/}({\it i\/}: integer): integer              \\
\> \> {\bf if} {\it ptr\/}[{\it i\/}]$\,<0$ {\bf return} {\it i}         \\
\> \> {\it ptr\/}[{\it i\/}] := {\it find\/}({\it ptr\/}[{\it i\/}])     \\
\> \> {\bf return} {\it ptr\/}[{\it i\/}]                                \\
\> {\bf end function}
\end{tabbing}
This function takes an integer argument, which is the label of a site, and
returns the label of the root site of the cluster to which that site
belongs.  The pointers are stored in an array {\it ptr,} which is negative
for all root nodes and contains the label of the site pointed to otherwise.
A version of this function in C is included in Appendix~A.  A non-recursive
implementation of the function is of course possible (as is the case with
all uses of recursion), and we give two such implementations in the
appendix for those who prefer this approach.  However, both are somewhat
more complicated than the recursive version, and with the benefit of a
modern optimizing compiler are found to offer little speed advantage.

\begin{figure}
\begin{center}
\psfig{figure=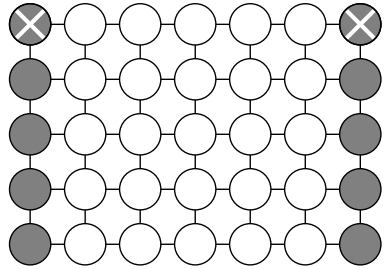,width=5cm}
\end{center}
\caption{Initial configuration of occupied (gray) and empty (white) sites
for checking for the presence of a spanning cluster.  In this example there
are no periodic boundary conditions on the lattice.  As the empty sites are
filled up during the course of the run, the two sites marked $\times$ will
have the same cluster root site if and only if there is a spanning cluster
on the lattice.}
\label{spanning}
\end{figure}

The program given in Appendix~A measures the largest cluster size as a
function of number of occupied sites.  Above the percolation transition in
the limit of large system size this quantity is equal to the size of the
giant component, and it also has interesting scaling properties below the
transition~\cite{Bazant99}.  However, there a many other quantities one
might want to measure using our algorithm.  Doing so usually involves
keeping track of some additional variables as the algorithm runs.  Here are
three examples.
\begin{enumerate}
\item {\bf Average cluster size:} In order to measure the average number of
  sites per cluster in site percolation it is sufficient to keep track of
  only the total number $c$ of clusters on the lattice.  Then the average
  cluster size is given by $n/c$, where $n$ is the number of occupied
  sites.  To keep track of $c$, we simply set it equal to zero initially,
  increase it by one every time a site is occupied, and decrease it by one
  every time we perform a union operation.  Similarly, for averages
  weighted by cluster size one can keep track of higher moments of the
  cluster-size distribution.
\item {\bf Cluster spanning:} In many calculations one would like to detect
  the onset of percolation in the system as sites or bonds are occupied.
  One way of doing this is to look for a cluster of occupied sites or bonds
  which spans the lattice from one side to the other.  Taking the example
  of site percolation, one can test for such a cluster by starting the
  lattice in the configuration shown in Fig.~\ref{spanning} in which there
  are occupied sites along two edges of the lattice and no periodic
  boundary conditions.  (Notice that the lattice is now not square.)  Then
  one proceeds to occupy the remaining sites of the lattice one by one in
  our standard fashion.  At any point, one can check for spanning simply by
  performing ``find'' operations on each of the two initial clusters,
  starting for example at the sites marked~$\times$.  If these two clusters
  have the same root site then spanning has occurred.  Otherwise it has
  not.
\item {\bf Cluster wrapping:} An alternative criterion for percolation is
  to use periodic boundary conditions and look for a cluster which wraps
  all the way around the lattice.  This condition is somewhat more
  difficult to detect than simple spanning, but with a little ingenuity it
  can be done, and the extra effort is, for some purposes, worthwhile.  An
  example is in the measurement of the position $p_c$ of the percolation
  transition.  As discussed in Section~\ref{critical}, estimates of $p_c$
  made using a cluster wrapping condition display significantly smaller
  finite-size errors than estimates made using cluster spanning on open
  systems.

\begin{figure}
\begin{center}
\psfig{figure=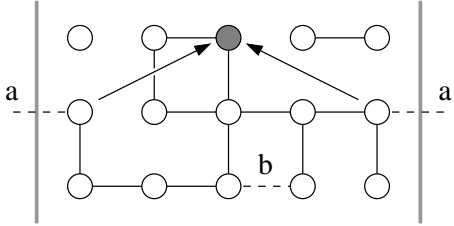,width=6cm}
\end{center}
\caption{Method for detecting cluster wrapping on periodic boundary
conditions.  When a bond is added~(a) which joins together two sites which
belong to the same cluster, it is possible, as here, that it causes the
cluster to wrap around the lattice.  To detect this, the displacements to
the root site of the cluster (shaded) are calculated (arrows).  If the
difference between these displacements is not equal to a single lattice
spacing, then wrapping has taken place.  Conversely if the bond (b) where
added, the displacements to the root site would differ by only a single
lattice spacing, indicating that wrapping has not taken place.}
\label{wrapping}
\end{figure}

A clever method for detecting cluster wrapping has been employed by
Machta~\etal~\cite{Machta96} in simulations of Potts models, and can be
adapted to the case of percolation in a straightforward manner.  We
describe the method for bond percolation, although it is easily applied to
site percolation also.  We add to each site two integer variables giving
the $x$- and $y$-displacements from that site to the site's parent in the
appropriate tree.  When we traverse the tree, we sum these displacements
along the path traversed to find the total displacement to the root site.
(We also update all displacements along the path when we carry out the path
compression.)  When an added bond connects together two sites which are
determined to belong to the same cluster, we compare the total
displacements to the root site for those two sites.  If these displacements
differ by just one lattice spacing, then cluster wrapping has not occurred.
If they differ by any other amount, it has.  This method is illustrated in
Fig.~\ref{wrapping}.
\end{enumerate}
It is worth noting also that, if one's object is only to detect the onset
of percolation, then one can halt the algorithm as soon as percolation is
detected.  There is no need to fill in any more sites or bonds once the
percolation point is reached.  This typically saves about 50\% on the
run-time of the algorithm, the critical occupation probability $p_c$ being
of the order of~$\half$.  A further small performance improvement can often
be achieved in this case by noting that it is no longer necessary to
generate a complete permutation of the sites (bonds) before starting the
algorithm.  In many cases it is sufficient simply to choose sites (bonds)
one by one at random as the algorithm progresses.  Sometimes in doing this
one will choose a site (bond) which is already occupied, in which case one
must generate a new random one.  The probability that this happens is equal
to the fraction $p$ of occupied sites (bonds), and hence the average number
of attempts we must make before we find an empty site is $1/(1-p)$.  The
total number of attempts made before we reach $p_c$ is therefore
\begin{equation}
N \int_0^{p_c} {\d p\over1-p} = -N\log(1-p_c),
\end{equation}
or $-M\log(1-p_c)$ for bond percolation.  If $p_c=0.5$, for example, this
means we will have to generate $N\log 2\simeq 0.693 N$ random numbers
during the course of the run, rather than the $N$ we would have had to
generate to make a complete permutation of the sites.  Thus it should be
quicker not to generate the complete permutation.  Only once $p_c$ becomes
large enough that $-\log(1-p_c)\gtrsim 1$ does it start to become
profitable to calculate the entire permutation, i.e.,~when
$p_c\gtrsim1-1/\e\simeq0.632$.  If one were to use a random selection
scheme for calculations over the whole range $0\le p\le1$, then the
algorithm would take time
\begin{equation}
N \int_0^{1-1/N} {\d p\over1-p} = N \log N
\end{equation}
to find and occupy all empty sites, which means overall operation would be
$\O(N\log N)$ not $\O(N)$, so generating the permutation is crucial in this
case to ensure that running time is~$\O(N)$.

One further slightly tricky point in the implementation of our scheme is
the performance of the convolution, Eq.~\eref{convolution}, of the results
of the algorithm with the binomial distribution.  Since the number of sites
or bonds on the lattice can easily be a million or more, direct evaluation
of the binomial coefficients using factorials is not possible.  And for
high-precision studies, such as the calculations presented in
Section~\ref{applications}, a Gaussian approximation to the binomial is not
sufficiently accurate.  Instead, therefore, we recommend the following
method of evaluation.  The binomial distribution, Eq.~\eref{binomial}, has
its largest value for given $N$ and $p$ when $n=n_{\rm max}=pN$.  We
arbitrarily set this value to~1.  (We will fix it in a moment.)  Now we
calculate $B(N,n,p)$ iteratively for all other $n$ from
\begin{equation}
B(N,n,p) = \biggl\lbrace \begin{array}{ll}
             B(N,n-1,p) {N-n+1\over n} {p\over1-p}  \quad & \mbox{$n>n_{\rm max}$}\\
             B(N,n+1,p) {n+1\over N-n} {1-p\over p} \quad & \mbox{$n<n_{\rm max}$.}\\
           \end{array}
\end{equation}
Then we calculate the normalization coefficient $C = \sum_n B(N,n,p)$ and
divide all the $B(N,n,p)$ by it, to correctly normalize the distribution.

\section{Applications}
\label{applications}
In this section we consider a number of applications of our algorithm to
open problems in percolation theory.  Not all of the results given here are
new---some have appeared previously in Refs.~\cite{CNSW00} and~\cite{NZ00}.
They are gathered here to give an idea of the types of problems for which
our algorithm is an appropriate tool.

\subsection{Measurement of the position of the percolation transition}
\label{critical}
Our algorithm is well suited to the measurement of critical properties of
percolation systems, such as position of the percolation transition and
critical exponents.  Our first example application is the calculation of
the position $p_c$ of the percolation threshold for site percolation on the
square lattice, a quantity for which we currently have no exact result.

There are a large number of possible methods for determining the position
of the percolation threshold numerically on a regular
lattice~\cite{RSK78,RSK80,Ziff92}, different methods being appropriate with
different algorithms.  As discussed in Ref.~\onlinecite{NZ00}, our
algorithm makes possible the use of a particularly attractive method based
on lattice wrapping probabilities, which has substantially smaller
finite-size scaling corrections than methods employed previously.

\begin{figure}
\begin{center}
\psfig{figure=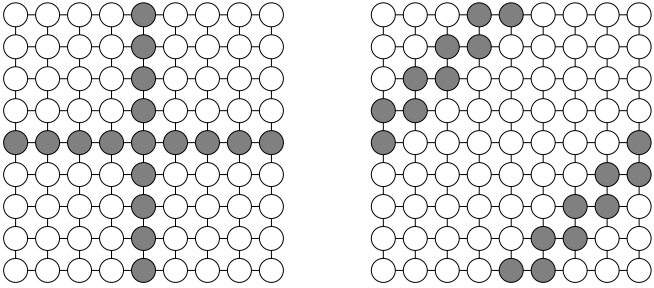,width=\columnwidth}
\end{center}
\caption{Two topologically distinct ways in which a percolation cluster can
  wrap around both axes of a two-dimensional periodic lattice.}
\label{both}
\end{figure}

We define $R_L(p)$ to be the probability that for site occupation
probability $p$ there exists a contiguous cluster of occupied sites which
wraps completely around a square lattice of $L\times L$ sites with periodic
boundary conditions.  As in Section~\ref{implementation}, cluster wrapping
is here taken to be an estimator of the presence or absence of percolation
on the infinite lattice.  There are a variety of possible ways in which
cluster wrapping can occur, giving rise to a variety of different
definitions for $R_L$:
\begin{itemize}
\item $R_L^{(h)}$ and $R_L^{(v)}$ are the probabilities that there exists a
  cluster which wraps around the boundary conditions in the horizontal and
  vertical directions respectively.  Clearly for square systems these two
  are equal.  In the rest of this paper we refer only to~$R_L^{(h)}$.
\item $R_L^{(e)}$ is the probability that there exists a cluster which
  wraps around either the horizontal or vertical directions, or both.
\item $R_L^{(b)}$ is the probability that there exists a cluster which
  wraps around both horizontal and vertical directions.  Notice that there
  are many topologically distinct ways in which this can occur.  Two of
  them, the ``cross'' configuration and the ``single spiral''
  configuration, are illustrated in Fig.~\ref{both}.  $R_L^{(b)}$~is
  defined to include both of these and all other possible ways of wrapping
  around both axes.
\item $R_L^{(1)}$ is the probability that a cluster exists which wraps
  around one specified axis but not the other axis.  As with $R_L^{(h)}$,
  it does not matter, for the square systems studied here, which axis we
  specify.
\end{itemize}
These four wrapping probabilities satisfy the equalities
\begin{eqnarray}
\label{equality1}
R_L^{(e)} &=& R_L^{(h)} + R_L^{(v)} - R_L^{(b)}
           =  2 R_L^{(h)} - R_L^{(b)},\\
\label{equality2}
R_L^{(1)} &=& R_L^{(h)} - R_L^{(b)} = R_L^{(e)} - R_L^{(h)}
           =  \mbox{$\frac12$} \bigl( R_L^{(e)} - R_L^{(b)} \bigr),
\end{eqnarray}
so that only two of them are independent.  They also satisfy the
inequalities $R_L^{(b)} \le R_L^{(h)} \le R_L^{(e)}$ and $R_L^{(1)} \le
R_L^{(h)}$.

Each run of our algorithm on the $L\times L$ square system gives one
estimate of each of these functions for the complete range of $p$.  It is a
crude estimate however: since an appropriate wrapping cluster either exists
or doesn't for all values of $p$, the corresponding $R_L(p)$ in the
microcanonical ensemble is simply a step function, except for $R_L^{(1)}$,
which has two steps, one up and one down.  All four functions can be
calculated in the microcanonical ensemble by finding the numbers
$n_c^{(h)}$ and $n_c^{(v)}$ of occupied sites at which clusters first
appear which wrap horizontally and vertically.  (This means that, as
discussed in Section~\ref{implementation}, the algorithm can be halted once
wrapping in both directions has occurred, which for the square lattice
gives a saving of about 40\% in the amount of CPU time used.)

Our estimates of the four functions are improved by averaging over many
runs of the algorithm, and the results are then convolved with the binomial
distribution, Eq.~\eref{convolution}, to give smooth curves of $R_L(p)$ in
the canonical ensemble.  (Alternatively, one can perform the convolution
first and average over runs second; both are linear operations, so order is
unimportant.  However, the convolution is quite a lengthy computation, so
it is sensible to choose the order that requires it to be performed only
once.)

\begin{figure}
\begin{center}
\psfig{figure=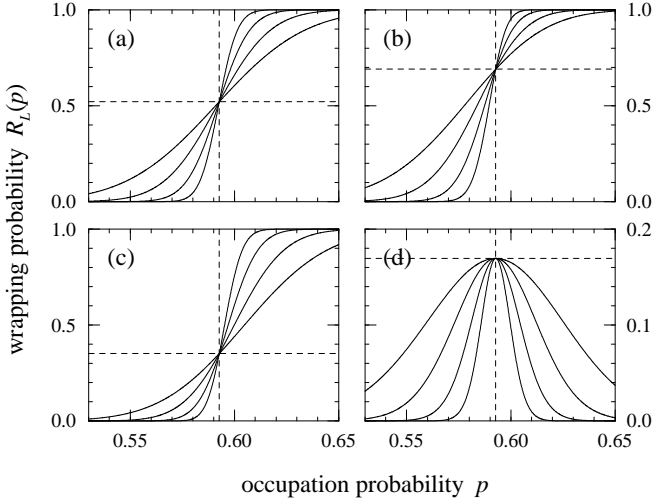,width=\columnwidth}
\end{center}
\caption{Plots of the cluster wrapping probability functions $R_L(p)$ for
  $L=32$, $64$, $128$ and $256$ in the region of the percolation transition
  for percolation (a)~along a specified axis, (b)~along either axis,
  (c)~along both axes, and (d)~along one axis but not the other.  Note that
  (d) has a vertical scale different from the other frames.  The dotted
  lines denote the expected values of $p_c$ and $R_\infty(p_c)$.}
\label{rl}
\end{figure}

In Fig.~\ref{rl} we show results from calculations of $R_L(p)$ using our
algorithm for the four different definitions, for systems of a variety of
sizes, in the vicinity of the percolation transition.

The reason for our interest in the wrapping probabilities is that their
values can be calculated exactly.  Exact expressions can be deduced from
the work of Pinson~\cite{Pinson94} and written directly in terms of the
Jacobi $\vartheta$-function $\vartheta_3(q)$ and the Dedekind
$\eta$-function~$\eta(q)$.  Pinson calculated the probability, which he
denoted $\pi(Z\times Z)$, of the occurrence on a square lattice of a
wrapping cluster with the ``cross'' topology shown in the left panel of
Fig.~\ref{both}.  By duality, our probability $R_\infty^{(e)}(p_c)$ is just
$1-\pi(Z \times Z)$, since if there is no wrapping around either axis, then
there must necessarily be a cross configuration on the dual lattice.  This
yields~\cite{ZLK99,note3}
\begin{eqnarray}
R_\infty^{(e)}(p_c) &=& 1 - \nonumber\\
& & \hspace{-14mm} {\vartheta_3(\e^{-3\pi/8}) \vartheta_3(\e^{-8\pi/3}) -
\vartheta_3(\e^{-3\pi/2}) \vartheta_3(\e^{-2\pi/3})
\over 2 [\eta(\e^{-2\pi})]^2 }.
\label{pinson1}
\end{eqnarray}
The probability $R_\infty^{(1)}(p_c)$ for wrapping around exactly one axis
is equal to the quantity denoted $\pi(1,0)$ by Pinson, which for a square
lattice can be written
\begin{equation}
R_\infty^{(1)}(p_c) =
{ 3\vartheta_3(\e^{-6\pi}) + \vartheta_3(\e^{-2\pi/3})-
4\vartheta_3(\e^{-8\pi/3}) \over\sqrt{6} [\eta(\e^{-2\pi})]^2 }.
\label{pinson2}
\end{equation}
The remaining two probabilities $R_\infty^{(h)}$ and $R_\infty^{(b)}$ can
now calculated from Eq.~\eref{equality2}.  To ten figures, the resulting
values for all four are:
\begin{eqnarray}
R_\infty^{(h)}(p_c) &=& 0.521058290,\quad
R_\infty^{(e)}(p_c) = 0.690473725,\nonumber\\
R_\infty^{(b)}(p_c) &=& 0.351642855,\quad
R_\infty^{(1)}(p_c) = 0.169415435.
\end{eqnarray}

If we calculate the solution $p$ of the equation
\begin{equation}
R_L(p) = R_\infty(p_c),
\label{estimator}
\end{equation}
we must have $p\to p_c$ as $L\to\infty$, and hence this solution is an
estimator for~$p_c$.  Furthermore, it is a very good estimator, as we now
demonstrate.

\begin{figure}
\begin{center}
\psfig{figure=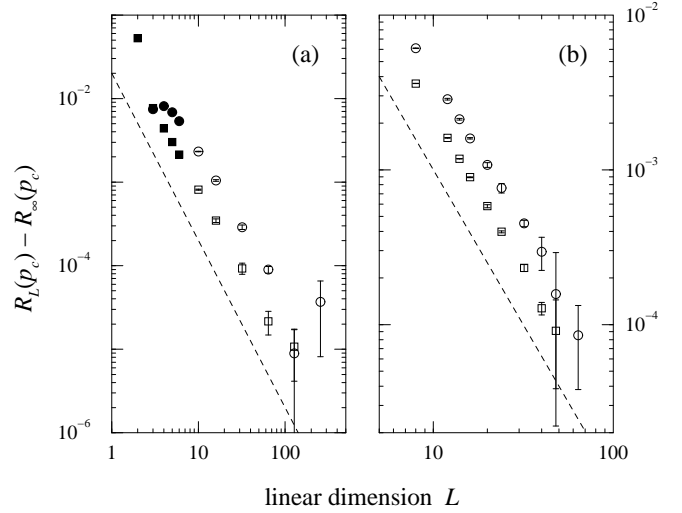,width=\columnwidth}
\end{center}
\caption{Convergence with increasing system size of $R_L(p_c)$ to its known
  value at $L=\infty$ for $R_L^{(e)}$ (circles) and $R_L^{(1)}$ (squares)
  under (a)~site percolation and (b)~bond percolation.  Filled symbols in
  panel~(a) are exact enumeration results for system sizes $L=2\ldots6$.
  The dotted lines show the expected slope if, as we conjecture, $R_L(p_c)$
  converges as $L^{-2}$ with increasing~$L$.  The data for $R_L^{(e)}$ in
  panel~(b) have been displaced upwards by a factor of two for clarity.}
\label{converge}
\end{figure}

In Fig.~\ref{converge}, we show numerical results for the finite-size
convergence of $R_L(p_c)$ to the $L=\infty$ values for both site and bond
percolation.  (For site percolation we used the current best-known value
for $p_c$ from Ref.~\onlinecite{NZ00}.)  Note that the expected statistical
error on $R_L(p_c)$ is known analytically to good accuracy, since each run
of the algorithm gives an independent estimate of $R_L$ which is either 1
or 0 depending on whether or not wrapping has occurred in the appropriate
fashion when a fraction $p_c$ of the sites are occupied.  Thus our estimate
of the mean of $R_L(p_c)$ over $n$ runs is drawn from a simple binomial
distribution which has standard deviation
\begin{equation}
\sigma_{R_L} = \sqrt{{R_L(p_c) [1-R_L(p_c)]\over n}}.
\label{rlerror}
\end{equation}
If we approximate $R_L(p_c)$ by the known value of $R_\infty(p_c)$, then we
can evaluate this expression for any~$n$.

As the figure shows, the finite-size corrections to $R_L$ decay
approximately as $L^{-2}$ with increasing system size.  For example, fits
to the data for $R_L^{(1)}$ (for which our numerical results are cleanest),
give slopes of $-1.95(17)$ (site percolation) and $-2.003(5)$ (bond
percolation).  On the basis of these results we conjecture that $R_L(p_c)$
converges to its $L=\infty$ value exactly as~$L^{-2}$.

At the same time, the width of the critical region is decreasing as
$L^{-1/\nu}$, so that the gradient of $R_L(p)$ in the critical region goes
as $L^{1/\nu}$.  Thus our estimate $p$ of the critical occupation
probability from Eq.~\eref{estimator} converges to $p_c$ according to
\begin{equation}
p - p_c \sim L^{-2-1/\nu} = L^{-11/4},
\end{equation}
where the last equality makes use of the known value $\nu=\frac43$ for
percolation on the square lattice.  This convergence is substantially
better than that for any other known estimator of~$p_c$.  The best
previously known convergence was $p-p_c\sim L^{-1-1/\nu}$ for certain
estimates based upon crossing probabilities in open systems, while many
other estimates, including the renormalization-group estimate, converge as
$p-p_c\sim L^{-1/\nu}$~\cite{Ziff92}.  It implies that we should be able to
derive accurate results for $p_c$ from simulations on quite small lattices.
Indeed we expect that statistical error will overwhelm the finite size
corrections at quite small system sizes, making larger lattices not only
unnecessary, but also essentially worthless.

Statistical errors for the calculation are estimated in conventional
fashion.  From Eq.~\eref{rlerror} we know that the error on the mean of
$R_L(p)$ over $n$ runs of the algorithm goes as $n^{-1/2}$, independent of
$L$, which implies that our estimate of $p_c$ carries an error
$\sigma_{p_c}$ scaling according to~$\sigma_{p_c}\sim n^{-1/2} L^{-1/\nu}$.
With each run taking time $\O(N)=\O(L^d)$, the total time $T\sim nL^d$
taken for the calculation is related to $\sigma_{p_c}$ according to
\begin{equation}
\sigma_{p_c} \sim {L^{d/2-1/\nu}\over\sqrt{T}} = {L^{1/4}\over\sqrt{T}},
\end{equation}
where the last equality holds in two dimensions.  Thus in order to make the
statistical errors on systems of different size the same, we should spend
an amount of time which scales as $T_L\sim L^{d-2/\nu}$ on systems of size
$L$, or $T_L\sim\sqrt{L}$ in two dimensions.

\begin{figure}
\begin{center}
\psfig{figure=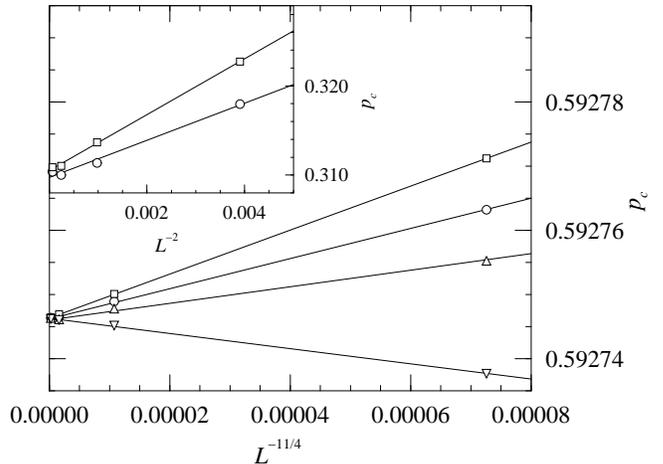,width=\columnwidth}
\end{center}
\caption{Finite size scaling of estimate for $p_c$ on square lattices of
$L\times L$ sites using measured probabilities of cluster wrapping along
one axis (circles), either axis (squares), both axes (upward-pointing
triangles), and one axis but not the other (downward-pointing triangles).
Inset: results of a similar calculation for cubic lattices of $L\times
L\times L$ sites using the probabilities of cluster wrapping along one axis
but not the other two (circles), and two axes but not the other one
(squares).}
\label{fss}
\end{figure}

In Fig.~\ref{fss} we show the results of a finite-size scaling calculation
of this type for $p_c$.  The four different definitions of $R_L$ give four
(non-independent) estimates of $p_c$: $0.59274621(13)$ for $R_L^{(h)}$,
$0.59274636(14)$ for $R_L^{(e)}$, $0.59274606(15)$ for $R_L^{(b)}$, and
$0.59274629(20)$ for $R_L^{(1)}$.  The first of these is the best, and is
indeed the most accurate current estimate of $p_c$ for site percolation on
the square lattice.  This calculation involved the simulation of more than
$7\times10^9$ separate samples, about half of which were for systems of
size $128\times128$.

The wrapping probability $R_L^{(1)}$ is of particular interest, because one
does not in fact need to know its value for the infinite system in order to
use it to estimate~$p_c$.  Since this function is non-monotonic it may
intercept its $L=\infty$ value at two different values of $p$, but its
maximum must necessarily converge to $p_c$ at least as fast as either of
these intercepts.  And the position of the maximum can be determined
without knowledge of the value of~$R_\infty^{(1)}(p_c)$.  In fact, in the
calculation shown in Fig.~\ref{fss}, we used the maximum of $R_L^{(1)}$ to
estimate $p_c$ and not the intercept, since in this case $R_L^{(1)}$ was
strictly lower than $R_\infty^{(1)}$ for all $p$, so that there were no
intercepts.

The criterion of a maximum in $R_L^{(1)}$ can be used to find the
percolation threshold in other systems for which exact results for the
wrapping probabilities are not known.  As an example, we show in the inset
of Fig.~\ref{fss} a finite-size scaling calculation of $p_c$ for
three-dimensional percolation on the cubic lattice (with periodic boundary
conditions in all directions) using this measure.  Here there are two
possible generalizations of our wrapping probability: $R_L^{(1)}(p)$ is the
probability that wrapping occurs along one axis and not the other two, and
$R_L^{(2)}(p)$ is the probability that wrapping occurs along two axes and
not the third.  We have calculated both in this case.

Although neither the exact value of $\nu$ nor the expected scaling of this
estimate of $p_c$ is known for the three-dimensional case, we can estimate
$p_c$ by varying the scaling exponent until an approximately straight line
is produced.  This procedure reveals that our estimate of $p_c$ scales
approximately as $L^{-2}$ in three dimensions, and we derive estimates of
$p_c=0.3097(3)$ and $0.3105(2)$ for the position of the transition.
Combining these results we estimate that $p_c=0.3101(10)$, which is in
reasonable agreement with the best known result for this quantity of
$0.3116080(4)$~\cite{LZ98,Ballesteros99}.  (Only a short run was performed
to obtain our result; better results could presumably be derived with
greater expenditure of CPU time.)

\subsection{Scaling of the wrapping probability functions}
It has been proposed~\cite{BW95,HA96} that below the percolation transition
the probability of finding a cluster which spans the system, or wraps
around it in the case of periodic boundary conditions, should scale
according to
\begin{equation}
R_L(p) \sim \exp(-L/\xi).
\label{stretched1}
\end{equation}
The probability of wrapping around the system is equal to the trace of the
product of the transfer matrices for the $L$ rows of the system.  With
periodic boundary conditions, the transfer matrices are equal for all rows,
and the wrapping probability is thus a simple sum of the $L$th powers of
the eigenvalues $\lambda_i$ of the individual transfer matrices: $R_L(p) =
\sum_i\lambda_i^L$.  For large $L$, this sum is dominated by the largest
eigenvalue $\lambda_0$ and $R_L(p) = \lambda_0^L = \exp(L\log\lambda_0)$.
Comparing with Eq.~\eref{stretched1}, we conclude that the leading constant
in~\eref{stretched1} must tend to unity as $L$ becomes large, and thus
\begin{equation}
R_L(p) = \exp(-L/\xi) = \exp[-cL(p_c-p)^\nu],
\label{stretched2}
\end{equation}
where $c$ is another constant.  In other words, as a function of $p_c-p$,
the wrapping probability should follow a stretched exponential with
exponent~$\nu$ and a leading constant of~1.  This contrasts with previous
conjectures that $R_L(p)$ has Gaussian tails~\cite{SA92,LSSE76,Wester00}.

\begin{figure}
\begin{center}
\psfig{figure=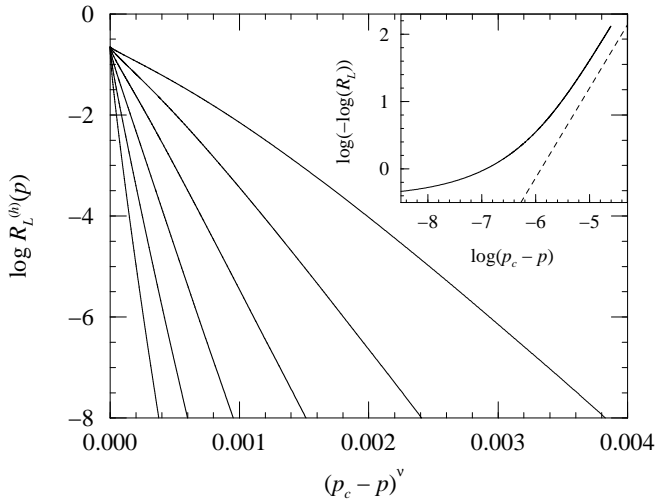,width=\columnwidth}
\end{center}
\caption{Demonstration of the stretched exponential behavior,
  Eq.~\eref{stretched2}, in simulation results for $1024\times1024$ square
  systems.  The results are averaged over $1\,000\,000$ runs of the
  algorithm.  The curves are for (top to bottom) $\nu=1.2$, 1.3, 1.4, 1.5,
  1.6, and 1.7.  Inset: the same data replotted logarithmically to show the
  stretched exponential behavior.  The dotted line indicates the expected
  slope of~$\frac43$.}
\label{stexp}
\end{figure}

The behavior~\eref{stretched2} is only seen when the correlation length is
substantially smaller than the system dimension, but also greater than the
lattice spacing, i.e.,~$1\ll\xi\ll L$.  This means that in order to observe
it clearly we need to perform simulations on reasonably large systems.  In
Fig.~\ref{stexp} we show results for site percolation on square lattices of
$1024\times1024$ sites, with $\log R_L$ plotted against $(p_c-p)^\nu$ for
various values of $\nu$, to look for straight-line behavior.
Interpretation of the results is a little difficult, since one must
discount curvature close to the origin where $\xi\gtrsim L$.  However, the
best straight line seems to occur in the region of $\nu=1.4\pm0.1$, in
agreement with the expected $\nu=\frac43$, while strongly ruling out the
Gaussian behavior.

A better demonstration of this result is shown in the inset of the figure.
Here we plot $\log(-\log(R_L))$ as a function of $\log(p_c-p)$, which,
since the leading constant in Eq.~\eref{stretched2} is equal to unity,
should give a straight line with slope $\frac43$ in the regime where
$1\ll\xi\ll L$.  This behavior is clearly visible in the figure.  Note that
this kind of plot is only valid for periodic boundary conditions, since the
leading constant in Eq.~\eref{stretched1} is not in general equal to~1 in
other cases.

\subsection{Percolation on random graphs}
For our last example, we demonstrate the use of our algorithm on a system
which is not built upon a regular lattice.  The calculations of this
section are instead performed on random graphs, i.e.,~collections of sites
(or ``vertices'') with random bonds (or ``edges'') between them.

Percolation can be considered a simple model for the robustness of
networks~\cite{CEBH00,CNSW00}.  In a communications network, messages are
routed from source to destination by passing them from one vertex to
another through the network.  In the simplest approximation, the network
between a given source and destination is functional so long as a single
path exists from source to destination, and non-functional otherwise.  In
real communications networks such as the Internet, a significant fraction
of the vertices (routers in the case of the Internet) are non-functional at
all times, and yet the majority of the network continues to function
because there are many possible paths from each source to each destination,
and it is rare that all such paths are simultaneously interrupted.  The
question therefore arises: what fraction of vertices must be non-functional
before communications are substantially disrupted?  This question may be
rephrased as a site percolation problem in which occupied vertices
represent functional routers and unoccupied vertices non-functional ones.
So long as there is a giant component of connected occupied vertices (the
equivalent of a percolating cluster on a regular lattice) then long-range
communication will be possible on the network.  Below the percolation
transition, where the giant component disappears, the network will fragment
into disconnected clusters.  Thus the percolation transition represents the
point at which the network as a whole becomes non-functional, and the size
of the giant component, if there is one, represents the fraction of the
network which can communicate effectively.

Both the position of the phase transition and the size of the giant
component have been calculated exactly by Callaway~\etal~\cite{CNSW00} for
random graphs with arbitrary degree distributions.  The degree $k$ of a
vertex in a network is the number of other vertices to which it is
connected.  If $p_k$ is the probability that a vertex has degree $k$ and
$q$ is the occupation probability for vertices, then the percolation
transition takes place at
\begin{equation}
q_c = {\sum_{k=0}^\infty k(k-1) p_k\over\sum_{k=0}^\infty kp_k},
\end{equation}
and the fraction $S$ of the graph filled by the giant component is the
solution of
\begin{equation}
S = q - q\sum_{k=0}^\infty p_k u^k,\qquad
u = 1 - q + q {\sum_{k=0}^\infty kp_k u^{k-1}\over\sum_{k=0}^\infty kp_k}.
\label{gcexact}
\end{equation}
For the Internet, the degree distribution has been found to be power-law in
form~\cite{FFF99}, though in practice the power law must have some cutoff
at finite $k$.  Thus a reasonable form for the degree distribution is
\begin{equation}
p_k = C k^{-\tau} \e^{-k/\kappa} \qquad \mbox{for $k\ge1$.}
\label{powerlaw}
\end{equation}
The exponent $\tau$ is found to take values between 2.1 and 2.5 depending
on the epoch in which the measurement was made and whether one looks at the
network at the router level or at the coarser domain level.  Vertices with
degree zero are excluded from the graph since a vertex with degree zero is
necessarily not a functional part of the network.

\begin{figure}
\begin{center}
\psfig{figure=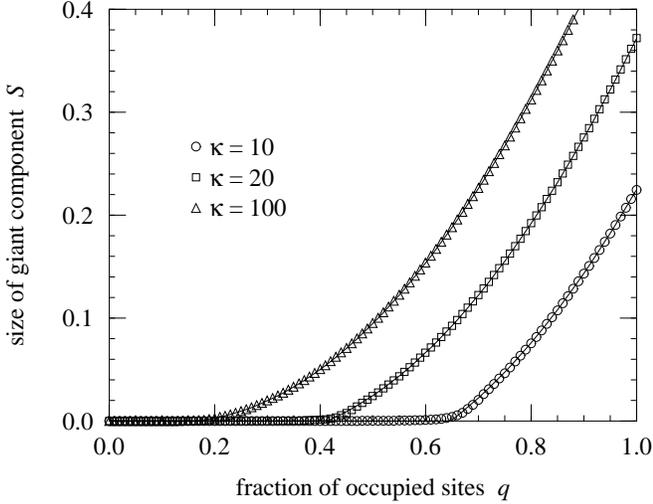,width=\columnwidth}
\end{center}
\caption{Simulation results (points) for site percolation on random graphs
  with degree distribution given by Eq.~\eref{powerlaw}, with $\tau=2.5$
  and three different values of $\kappa$.  The solid lines are the exact
  solution of Callaway~\etal~\protect\cite{CNSW00}, Eq.~\eref{gcexact}, for
  the same parameter values.}
\label{rggc}
\end{figure}

We can generate a random graph of $N$ vertices with this degree
distribution in $\O(N)$ time using the prescription given in
Ref.~\onlinecite{NSW00} and then use our percolation algorithm to
calculate, for example, the size of the largest cluster for all values
of~$q$.  In Fig.~\ref{rggc} we show the results of such a calculation for
$N=1\,000\,000$ and $\tau=2.5$ for three different values of the cutoff
parameter~$\kappa$, along with the exact solution derived by numerical
iteration of Eq.~\eref{gcexact}.  As the figure shows, the two are in
excellent agreement.  The simulations for this figure took about an hour in
total.  We would expect a simulation performed using the standard
depth-first search and giving results of similar accuracy to take about a
million times as long, or about a century.

A number of authors~\cite{CNSW00,AJB00,Broder00} have examined the
resilience of networks to the selective removal of the vertices with
highest degree.  This scenario can also be simulated efficiently using our
algorithm.  The only modification necessary is that the vertices are now
occupied in order of increasing degree, rather than in random order as in
the previous case.  We note however that the average time taken to sort the
vertices in order of increasing degree scales as $\O(N\log N)$ when using
standard sorting algorithms such as quicksort~\cite{Sedgewick88}, and hence
this calculation is dominated by the time to perform the sort for large
$N$, making overall running time $\O(N\log N)$ rather than $\O(N)$.

\section{Conclusions}
\label{concs}
We have described in detail a new algorithm for studying site or bond
percolation on any lattice which can calculate the value of an observable
quantity for all values of the site or bond occupation probability from
zero to one in time which, for all practical purposes, scales linearly with
lattice volume.  We have presented a time complexity analysis demonstrating
this scaling, empirical results showing the scaling and comparing running
time to other algorithms for percolation problems, and a description of the
details of implementation of the algorithm.  A full working program is
presented in the following appendix and is also available for download.  We
have given three example applications for our algorithm: the measurement of
the position of the percolation transition for site percolation on a square
lattice, for which we derive the most accurate result yet for this
quantity; the confirmation of the expected $\frac43$-power stretched
exponential behavior in the tails of the wrapping probability functions for
percolating clusters on the square lattice; and the calculation of the size
of the giant component for site percolation on a random graph, which
confirms the recently published exact solution for the same quantity.

\section{Acknowledgements}
The authors would like to thank Cris Moore, Barak Pearlmutter, and Dietrich
Stauffer for helpful comments.  This work was supported in part by the
National Science Foundation and Intel Corporation.

{
\appendix
\renewcommand{\baselinestretch}{1.0}
\section{Program}
In this appendix we give a complete program in C for our algorithm for site
percolation on a square lattice of $N=L\times L$ sites with periodic
boundary conditions.  This program prints out the size of the largest
cluster on the lattice as a function of number of occupied sites $n$ for
values of $n$ from 1 to $N$.  The entire program consists of 73 lines of
code.

First we set up some constants and global variables:
\begin{verbatim}
#include <stdlib.h>
#include <stdio.h>

#define L 128        /* Linear dimension */
#define N (L*L)
#define EMPTY (-N-1)

int ptr[N];          /* Array of pointers */
int nn[N][4];        /* Nearest neighbors */
int order[N];        /* Occupation order */
\end{verbatim}
Sites are indexed with a single signed integer label for speed, taking
values from 0 to $N-1$.  Note that on computers which represent integers in
32 bits, this program can, for this reason, only be used for lattices of up
to $2^{31}\simeq2$~billion sites.  While this is adequate for most
purposes, longer labels will be needed if you wish to study larger
lattices.

The array {\tt ptr[]} serves triple duty: for non-root occupied sites it
contains the label for the site's parent in the tree (the ``pointer'');
root sites are recognized by a negative value of {\tt ptr[]}, and that
value is equal to minus the size of the cluster; for unoccupied sites {\tt
ptr[]} takes the value {\tt EMPTY}.

Next we set up the array {\tt nn[][]} which contains a list of the nearest
neighbors of each site.  Only this array need be changed in order for the
program to work with a lattice of different topology.
\begin{verbatim}
void boundaries()
{
  int i;

  for (i=0; i<N; i++) {
    nn[i][0] = (i+1)%N;
    nn[i][1] = (i+N-1)%N;
    nn[i][2] = (i+L)%N;
    nn[i][3] = (i+N-L)%N;
    if (i%L==0) nn[i][1] = i+L-1;
    if ((i+1)%L==0) nn[i][0] = i-L+1;
  }
}
\end{verbatim}
Now we generate the random order in which the sites will be occupied, by
randomly permuting the integers from 0 to $N-1$:
\begin{verbatim}
void permutation()
{
  int i,j;
  int temp;

  for (i=0; i<N; i++) order[i] = i;
  for (i=0; i<N; i++) {
    j = i + (N-i)*drand();
    temp = order[i];
    order[i] = order[j];
    order[j] = temp;
  }
}
\end{verbatim}
Here the function {\tt drand()} generates a random double precision
floating point number between 0 and 1.  Many people will have such a
function already to hand.  For those who don't, a suitable one is supplied
with Ref.~\cite{download}.

We also define a function which performs the ``find'' operation, returning
the label of the root site of a cluster, as well as path compression.  The
version we use is recursive, as described in Section~\ref{implementation}:
\begin{verbatim}
int findroot(int i)
{
  if (ptr[i]<0) return i;
  return ptr[i] = findroot(ptr[i]);
}
\end{verbatim}
The code to perform the actual algorithm is quite brief.  It works exactly
as described in the text.  Sites are occupied in the order specified by the
array {\tt order[]}.  The function {\tt findroot()} is called to find the
roots of each of the adjacent sites.  If amalgamation is needed, it is
performed in a weighted fashion, smaller clusters being added to larger
(bearing in mind that the value of {\tt ptr[]} for the root nodes is {\em
minus\/} the size of the corresponding cluster).
\begin{verbatim}
void percolate()
{
  int i,j;
  int s1,s2;
  int r1,r2;
  int big=0;

  for (i=0; i<N; i++) ptr[i] = EMPTY;
  for (i=0; i<N; i++) {
    r1 = s1 = order[i];
    ptr[s1] = -1;
    for (j=0; j<4; j++) {
      s2 = nn[s1][j];
      if (ptr[s2]!=EMPTY) {
        r2 = findroot(s2);
        if (r2!=r1) {
          if (ptr[r1]>ptr[r2]) {
            ptr[r2] += ptr[r1];
            ptr[r1] = r2;
            r1 = r2;
          } else {
            ptr[r1] += ptr[r2];
            ptr[r2] = r1;
          }
          if (-ptr[r1]>big) big = -ptr[r1];
        }
      }
    }
    printf("%i %i\n",i+1,big);
  }
}
\end{verbatim}
The main program is now simple:
\begin{verbatim}
main()
{
  boundaries();
  permutation();
  percolate();
}
\end{verbatim}
A complete working version of this program can also be downloaded from the
Internet~\cite{download}.

While our recursive implementation of the function \verb|findroot()| is
concise, some readers may find it unsatisfactory, either because they are
using a compiler under which recursive code runs slowly, or because they
want to translate the program into another language, such as Fortran~77,
which does not support recursion.  For their benefit we give here two
alternative implementations of this function, neither of which makes use of
recursion.  The first of these is a straightforward implementation
combining the find operation with path compression, as before, but using an
explicit stack:
\begin{verbatim}
#define STACKSIZE 100

int findroot(int i)
{
  int r;
  int sp=0;
  int stack[STACKSIZE];

  r = i;
  while (ptr[r]>=0) {
    stack[sp++] = r;
    r = ptr[r];
  }
  while (sp) ptr[stack[--sp]] = r;
  return r;
}
\end{verbatim}
The stack used is small, having just 100 elements.  This should be more
than sufficient in almost all cases, since the average distance traversed
across the tree is only about~3.

A more elegant way to implement \verb|findroot()| without recursion is to
modify the union/find algorithm itself slightly.  There is, it turns out,
another union/find algorithm which runs in $\O(N)$ time.  In this algorithm
the union operation is as before, but the find operation now involves
``path halving'' instead of path compression.  With path halving, each
pointer along the path traversed is changed to point to its ``grandparent''
in the tree, which effectively halves the length of the path from a site to
the root of the cluster each time \verb|findroot()| is called.
Tarjan~\cite{Tarjan83} has shown that this find operation also runs
asymptotically in very nearly constant time, giving an algorithm which runs
in linear time overall.  Here is a version of the function
\verb|findroot()| which implements path halving:
\begin{verbatim}
int findroot(int i)
{
  int r,s;

  r = s = i;
  while (ptr[r]>=0) {
    ptr[s] = ptr[r];
    s = r;
    r = ptr[r];
  }
  return r;
}
\end{verbatim}
}


\begin{references}
\bibitem{note2}
We did not, as the reader may have guessed, actually run the depth-first
search algorithm for 4.5 million seconds (about 2 months), since this would
have been wasteful of computing resources.  Instead, we ran the algorithm
for 100 representative values of $n$, the number of occupied bonds, and
scaled the resulting timing up to estimate the run-time for all
$1\,000\,000$ values.
\end{references}
\end{document}